\documentclass[prb,preprint,amsmath,amssymb,showpacs]{revtex4-1}
\usepackage{graphicx,psfrag,color,hyperref}
\usepackage{amssymb,latexsym,mathrsfs,amsmath,mathrsfs}
\usepackage{float}
\usepackage{array}
\usepackage{setspace}
\usepackage{slashbox}
\usepackage{diagbox}
\usepackage{makecell}
\usepackage{soul}
\usepackage{setspace}
\usepackage{mwe}
\usepackage{adjustbox}
\usepackage{float}
\usepackage{multirow}
\usepackage{chngpage}

\begin{document}

\title{Quantitative phase analysis of Bi$_{2}$Sr$_{2}$CaCu$_{2}$O$_{8+x}$ and competing intergrowth and co-crystallising phases via a Rietveld refinement study.}
	
\author{Neeraj K. Rajak}
\affiliation{School of Physics, IISER Thiruvananthapuram, Vithura, Kerala-695551, India}
\author{Arya M.}
\affiliation{School of Physics, IISER Thiruvananthapuram, Vithura, Kerala-695551, India}
\author{D. Jaiswal-Nagar*}
\affiliation{School of Physics, IISER Thiruvananthapuram, Vithura, Kerala-695551, India}
*Email:deepshikha@iisertvm.ac.in	

\begin{abstract}
We have used Rietveld refinement technique to determine the extent of intergrowth of Bi$_{2}$Sr$_{2}$CuO$_{6+x}$ phase and co-crystallization of competing phases in a high temperature superconductor Bi$_{2}$Sr$_{2}$CaCu$_{2}$O$_{8+x}$. The refinement was done on powder diffractograms obtained on powders of Bi$_{2}$Sr$_{2}$CaCu$_{2}$O$_{8+x}$ made by grinding single crystals of Bi$_{2}$Sr$_{2}$CaCu$_{2}$O$_{8+x}$ grown using two different self-flux techniques, namely pressure technique and regrowth technique. JANA and FullPROF Rietveld programs were used for refinement, both of which gave consistent results. Bi and Sr atom's positions were refined in the average structure of centrosymmetric space group $Bbmb$. To incorporate for Bi atoms modulation and extract information about the modulation vector, refinement was done in centrosymmetric space group N$^{Bbmb}_{1\bar{1}1}$(Bbmb(0$\gamma$1)). It was found that the b* component of the modulation vector decreased with a decrease in the superconducting transition temperature in pressure technique sample compared to the regrowth sample, suggesting a better alignment of the CuO$_2$ planes with respect to Bi-O planes in pressure technique sample. All the samples were also found to exhibit strong preferred orientation effect. Values of March-Dollase parameters corresponding to the preferred orientation function were obtained. We also calculated Brindley absorption contrast factors \textit{t}$_{\phi}$ and the effect of micro-absorption on the quantity of phases present in each sample. A Rietveld refinement incorporating all the factors resulted in exceptional values of goodness of fit parameters for all the samples with the lowest value of 2.08 found for the pressure technique sample ground for 2 minutes. Additionally, the powders corresponding to the pressure technique crystals were found to have no co-crystallising phase and $\sim$ 94$\%$ Bi-2212 phase, suggesting that crystals grown by pressure technique are of extremely good quality, much better than those grown by regrowth flux technique. 
\end{abstract}

\maketitle

\section{Introduction} 
~~~The Bi-Sr-Ca-Cu-O (BSCCO) system consists of three superconducting phases, Bi$_{2}$Sr$_{2}$CuO$_{6+x}$ (Bi-2201) having a T$_c$ $\sim$ 20 K, Bi$_{2}$Sr$_{2}$CaCu$_{2}$O$_{8+x}$ (Bi-2212) with a T$_c$ $\sim$ 90 K and Bi$_{2}$Sr$_{2}$Ca$_{2}$Cu$_{3}$O$_{10+x}$ (Bi-2223) with a T$_c$ $\sim$ 110 K \cite{majewski,majewski_matter_res_soc,majewsky_mr,maltsev,styve,majewski_adv_matter,wesche,kharissova,muller,schulze,hiraga}. They can be represented by the general formula Bi$_{2}$Sr$_{2}$Ca$_{n-1}$Cu$_{n}$O$_{2n+4+x}$, where n = 1,2,3.... "n" represents the number of CuO$_2$ layers in the structure. So, 2201 phase has 1 CuO$_2$ layer; 2212 phase has 2 CuO$_2$ and 2223 phase has 3 CuO$_2$ layers. The "n" phase arises due to diffusion driven insertion of Cu-O and Ca planes within the (n-1) phase. This, often, results in an incomplete crystallisation of the "n" phase such that the "n" phase contains unit cells of the (n-1) or the (n-2) phase, usually referred to as intergrowths \cite{manaila,malis}. Amongst the three phases of the BSCCO system, Bi-2212 is known to have the largest domain of stability in the temperature-composition phase space \cite{majewski,muller,schulze}, thus ensuring, that the 2212 phase is the easiest to grow in the BSCCO system. Because of the presence of many stable phases in the same range of pressures and compositions as the BSCCO system, a large number of other competing phases co-cyrstallise in the multi-component Bi$_2$O$_3$-SrO-CaO-CuO system, seriously affecting phase pure crystal growth of BSCCO system \cite{maltsev,majewski,wesche,kharissova}. The most prominent impurity phase in the Bi-2212 system is the intergrowth of Bi-2201 phase \cite{maltsev}, a homologue of Bi-2212 with no Calcium atom in the unit cell. Impurity phases other than Bi-2201 which are known to commonly co-crystallize with Bi-2212 are (SrCa)CuO$_2$, (CaSr)Al$_2$O$_5$, Ca$_2$CuO$_3$, Sr$_{14-x}$Ca$_x$Cu$_{24}$O$_{41-y}$ (x=7), Sr$_{3-x}$Ca$_x$Cu$_{24}$O$_{41-y}$ (x = 1) \cite{majewski,muller,schulze}. Since the superconducting properties of the main phase may be seriously affected by the presence of the intergrowth and co-crystallising phases, it is necessary to quantify the extent of co-crystallization and inter-growth in the BSCCO system.\\
A large number of techniques exist in the literature that use intensities obtained from powder x-ray diffraction (PXRD) for a quantitative phase analysis (QPA). Examples include the internal standard method \cite{alexander,luo,luo1}, matrix-flushing technique \cite{chung,zhu}, stacking fault defect technique \cite{malis,manaila} etc. However, each technique suffers from its own problem, for example, internal standard method involves constructing cumbersome calibration curve of each component in a mixture from standards where the standard contains exactly the same percentage of pure reference material \cite{alexander,luo,luo1} etc. Other techniques like transmission electron microscopy, high-resolution transmission electron microscopy, that could be used to ascertain impurity phases in a material suffer from the problem of them being local and thickness resoluted probes. Additionally, the particular case of BSCCO-2212 that has the intergrowth phase of BSCCO-2201 in the c-direction, requires complicated and expensive sample preparation technique of ``focussed ion-beam milling" \cite{hiraga}. However, even this specialised measurement does not have the capability of giving very detailed information regarding structure, mismatch of planes with respect to each other, modulation of atoms and their correlation etc. to a quantitative estimation of the amount of phases present in a batch of crystals. Rietveld refinement technique \cite{carvajal,carvajal1,rietveld,rietveld1}, in contrast, is a very powerful technique that has the capability of providing an accurate and reliable means of QPA using a powder diffractogram where the whole powder diffraction pattern of a given specimen is completely fitted, on a point by point basis, by incorporating effects such as atoms position, its occupancy and modulation, preferred orientation, x-ray absorption, micro-absorption etc. to obtain parameters defining the individual Bragg reflections. Since the Rietveld technique includes these parameters in obtaining a fit to the powder diffractogram, this technique has the additional advantage of providing information on not only how the intrinsic structure of a given phase affects the final phase fraction estimation but also how external parameters like orientation of the grains themselves affect the phase estimation of a given component quantitatively.\\ 
In this work, we describe detailed Rietveld refinement study of powders of Bi-2212 that were made by powdering single crystals of Bi-2212 grown using self-flux techniques. In order to see the effect of grinding time on the phase estimation of Bi-2212, since a large grinding time results in breakage of Bi-O layers in Bi-2212 superconductor, we ground the powders in two limits: (1) a high grinding time of 2 hour (RG1-2hour, RG2-2hour and PT-2hour samples) and (2) a low grinding time of 2 minutes (RG1-2min, RG2-2min and PT-2min samples). The refinement was done using JANA and FullPROF softwares. The average structure of the main phase of Bi-2212 was refined in the centrosymmetric space group $Bbmb$, while the modulation of the Bi atom was refined in the centrosymmetric space group N$^{Bbmb}_{1\bar{1}1}$(Bbmb(0$\gamma$1). Difference Fourier maps of the modulated Bi1 atom was drawn for each of the 6 powder diffractogram. Average structure and modulation of Bi atom's refinement resulted in a decrease of the b* component of the modulation vector with a decrease in superconducting transition temperature for PT samples compared to the RG samples, pointing to better quality of pressure technique crystals. Since the powders were made by crushing single crystals, a large preferred orientation was found to exist in the powders that was accounted for by refining the preferred orientation function. Absorption correction was incorporated in all the 6 powders. The quantitative phase analysis showed that pressure technique samples to have $\sim$ 94$\%$ of the main Bi-2212 phase suggesting that the parent single crystals from which the powders were made are of extremely good quality. It was also found that the phase estimation is effectively independent of the grinding time in PT powders, with slight difference in phase estimation for RG1 samples. However, the grinding time seemed to have a large effect on RG2 samples, depending on the presence or absence of a new phase, Phase$\#$3, that we discovered.  
\section{Experimental Details}
The powder diffractograms that were subjected to a Rietveld analysis were obtained by crushing the single crystals of Bi-2212 grown using two different self-flux techniques, namely regrowth technique and pressure technique \cite{rajak}. In the regrowth technique, the crystals are first obtained using the self-flux method. Crystals, so obtained, were nomenclatured as RG1. In the second step of the regrowth method, RG1 crystals are ground and allowed to crystallise again using the same temperature-time profile as that used in the first crystallisation \cite{rajak}. Single crystals obtained from the second step of regrowth technique were called RG2. In the pressure technique, Bi-2212 crystals were obtained using a small pressure of 0.234 N/m$^2$. These crystals were named PT \cite{rajak}. In order to get the powders for a QPA, crystals obtained by a given technique and grown in the same batch, were ground into a fine powder manually using agate mortar and pestle, for the same time of 2 hours. We also made powders by similar grinding but for a much lesser time of 2 minutes, for reasons explained in sub-section "Background correction". The former powders are named as RG1-2hour, RG2-2hour and PT-2hour while the latter are named as RG1-2min, RG2-2min and PT-2min. The powders were then mounted on PANalytical's zero background sample holder (model number SI SUBSTER 32 MM), and pressed gently with a dried glass slide. The powder diffractograms were recorded on a PANalytical Bragg-Brentano geometry powder diffractometer with copper source, having both Cu-K$\alpha$$_{1}$ and Cu-K$\alpha$$_{2}$ radiations, corresponding to wavelengths of 1.540$\mathring{A}$ and 1.544$\mathring{A}$ respectively. The counts were recorded for a total time of 2 hours at intervals of 0.016$^{\circ}$ in the 2$\theta$ range of 5–90$^{\circ}$. Rietveld analysis of the full diffraction profile was done using the FullPROF program -  FullPROF.2k (Version 6.20 - Jan2018-ILL JRC)  \cite{carvajal,carvajal1} as well as JANA program - Jana2006 Version : 10/10/2019 \cite{petricekjana}. For a QPA, a knowledge of the grain-size distribution of the powders is necessary; this was determined using Beckman Coulter LS12 320 laser diffraction particle size analyser.
\section{Results and analysis} 
In the Rietveld method, the calculation of the diffraction profile needs the unit-cell parameters of the contributing phases as one of the inputs. In the present analysis, we have incorporated the major phase Bi-2212, the intergrowth phase Bi-2201 as well as a newly discovered impurity phase that we call Phase$\#$3 \cite{rajakcmcm}. The initial data for the Bi-2212 phase was taken from Patrick et al. \cite{petricek} and Akhtar et al. for the Bi2201 phase \cite{akhtar}. The only structure parameters of the 2201 phase which was refined were the cell constants. The structure of the new phase that we identified was found to crystallise in the orthorhombic space group Cmcm. The details of the structure will be reported elsewhere \cite{rajakcmcm}. The refinement was started with a background correction that consisted of linear interpolation of manually selected background data points, selected graphically, from the diffraction data. The scale factors and the zero shifts were, then, refined while keeping the structure factors normalized \cite{carvajal,carvajal1}. The obtained values for instrumental zero correction for all the powders is tabulated in Table \ref{table:final fit values}. This was followed by a refinement of the unit cell length parameters $a$, $b$ and $c$. The angles $\alpha$, $\beta$ and $\gamma$ were all taken to be equal to 90$^{\circ}$ and were not refined. Finally, the profile parameters were refined, wherein, the parameters governing the instrumental contribution to the FWHM of the diffraction peaks were refined using the Caglioti function \cite{caglioti} obtained using a standard Y$_{2}$O$_{3}$ powder \cite{manjuBZO}. The profile parameters were refined using the pseudo-Voigt function given by equation \ref{eq:pseuo-voigt}:
\begin{center}
	\begin{equation}
	\label{eq:pseuo-voigt}
	\Omega(x) = \eta L(x) + (1-\eta) G(x);~~~~ \eta = \eta_0 + X 2\theta~~~~~~~ 0 \leq \eta \leq 1
	\end{equation}
\end{center}
where $L(x)$, $G(x)$ and $\eta$ are the Lorenzian, Gaussian and mixing parameters respectively. $\eta_0$ and $X$ are constants (to be obtained with a least square refinement) which determine the 2$\theta$ dependence of the mixing parameter $\eta$. The set of parameters used (in this work) for profile fit within the FullPROF and JANA software are different. In the FullPROF package, the set of refined parameters are U, V, W, X and $\eta_0$ such that the Lorentzian and the Gaussian width have the same value \textit{H} (see Table \ref{table:FullPROFVsJANA}). In the JANA Rietveld program, a modified implementation of pseudo-Voigt function by Thompson et al. \cite{thompson} (also known as TCH Voigt function) is used, wherein, the Gaussian width $H_G$ is modeled using the Caglioti function \cite{caglioti} while the Lorentzian term $H_L$ is modeled using Lorentzian isotropic crystalline size broadening term $L_x$ and isotropic strain broadening term $L_y$ as defined in Table \ref{table:FullPROFVsJANA}. The width \textit{H} is then calculated (see Table \ref{table:FullPROFVsJANA} coloum 2, row 6) as in \cite{thompson}. It was also noted that Caglioti parameters in JANA are 1000/(8*ln2) times that of FullPROF.\\
A selection of these parameters is, then, refined by minimising Goodness of fit, GOF \cite{toby}. The refinement starts out with a large value of GOF and parameters of the model are varied in such a manner that GOF is minimized as the fit improves.
\begin{table}[H]
	\centering
	\begin{adjustwidth}{-0.5in}{0.2in}
		\begin{tabular}{|c|c|c|}
			\hline 
			\makecell{Profile\\ function} &\multicolumn{2}{c|}{$\Omega$ = $\eta$L($H$,$\eta$) + (1-$\eta$)G($H$,$\eta$)} \\ \hline
			
			Program & FullPROF & JANA \\ \hline
			
			\makecell{Refined\\ parameters}& U,V,W,X,$\eta_0$ & GU,GV,GW,$L_x$,$L_y$ \\ \hline
			
			\makecell{Gaussian\\ width}& $\frac{H_G}{H}$ $=$ 1- 0.7441$\eta$ - 0.24781$\eta^2$ - 0.0081$\eta^3$ & $H_G^2$ $=$ GU$tan^2$$\theta$ + GV$tan$$\theta$ + GW \\ \hline
			
			\makecell{Lorentzian\\ width}& $\frac{H_L}{H}$ $=$ 0.72928$\eta$ + 0.19289$\eta^2$ + 0.07783 & $H_L$ $=$ $L_x$$tan\theta$ + $\frac{L_y}{tan\theta}$ \\ \hline
			
			\makecell{Pseudo\\ voigt \\ Width,$H$} & $H^2$ $=$ U$tan^2$$\theta$ + V$tan$$\theta$ + W & \makecell{$H^5$ $=$ $H_G^5$ + $2.69269 H_G^4 H_L$ + $2.42843 H_G^3 H_L^2$ + \\$4.47163 H_G^2 H_L^3$ + $0.07842 H_G H_L^4$ + $H_L^5$} \\ \hline
			
			\makecell{Voigt\\mixing\\parameter}& $\eta$ = $\eta_0$ + $X$2$\theta$ & $\eta$ = 1.36603($\frac{H_L}{H}$) - 0.47719($\frac{H_L}{H}$)$^2$ + 0.11116($\frac{H_L}{H}$)$^3$ \\ \hline
		\end{tabular} 
	\end{adjustwidth}
	\caption{Profile functions used in JANA and FullPROF }
	\label{table:FullPROFVsJANA}
\end{table}
In order to fit the sample profile, we refined the profile parameters $\eta_0$ and \textit{X} obtained from the instrument such that $\eta_0$ defined in equation \ref{eq:pseuo-voigt} remains within the bound 0 $<$ $\eta_0$ $<$ 1 for the angular range of 5$^{\circ} <$ 2$\theta < $ 80$^{\circ}$, for reasons explained in sections below. To account for the large difference between the calculated and the observed intensity, we refined the preferred orientation function, P$_{\phi,h}$, proposed by March and Dollase \cite{dollase}, since the polycrystalline grains of Bi-2212 are known to have substantial preferred orientation along $\{00l\}$. It is known that the main phase Bi-2212 and the impurity phase Bi-2201 have a considerable difference in their pure phase linear absorption coefficients ($\approx$ 200 $cm^{-1}$), so it was imperative to incorporate the absorption correction parameter A$_{\phi,h}$, whose variation with a variation in the average particle size of the intergrowth phase Bi-2201 was also calculated. Since, the new impurity phase Phase$\#$3's absorption coefficient is not known, A$_{\phi,h}$ was not incorporated for the Phase$\#$3. The sub-sections below describe the details of each parameter that was refined.
\begin{figure}
	\begin{center}
		\includegraphics[width=0.8\textwidth]{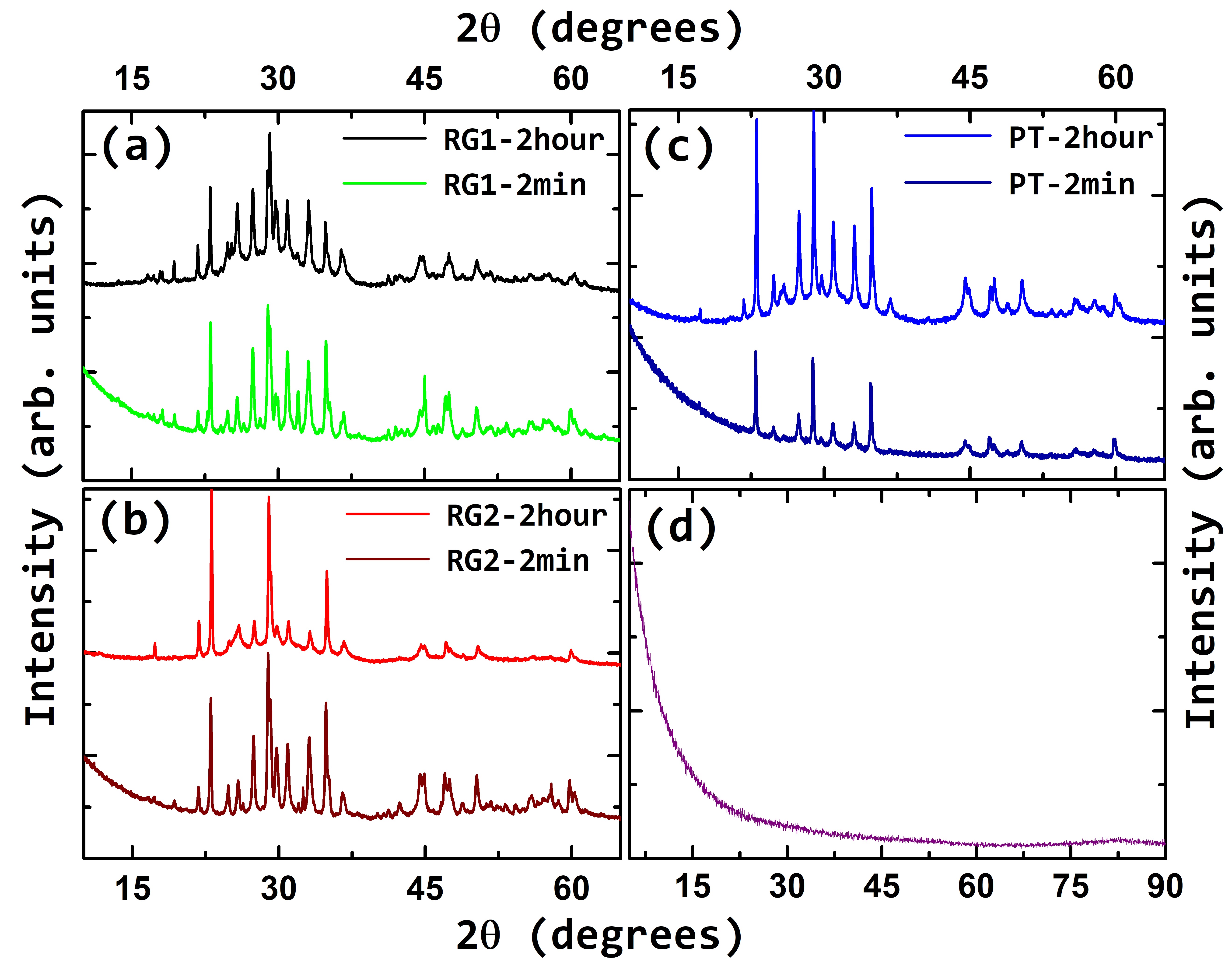}
		\caption{(colour online) (a)-(c): X-ray diffraction data obtained on crushed single crystals for two different grinding times of 2 minutes and 2 hours. Top graph in each panel represent the powder diffractogram taken after a grinding time of 2 hours while the lower graphs denote the 2 minute grinding time diffraction data. (a) RG1-2 min (green) and RG1-2hour (black). (b) RG2-2min (wine) and RG2-2hour (red) (c) PT-2min (dark blue) and PT-2hour (light blue). (d) Powder diffractogram acquired for 20 minutes with only the sample holder to account for sample holder's contribution to the background.}
		\label{fig:background}
	\end{center}
\end{figure}
\subsection{Background correction}
For a QPA, it is recommended that the powder must be prepared in such a way that it is homogeneous and there are sufficient number of particles with random orientation \cite{carvajal,carvajal1}. This condition is fulfilled if the powder is made after grinding for a sufficiently long time. So, black, red and light-blue curves in Figs. \ref{fig:background} (a)-(c) respectively represent the powder diffractograms of RG1, RG2 and PT powders respectively, obtained for a grinding time of 2 hours. It was found that the background for all these curves show a very similar trend with respect to 2$\theta$, namely, a very high value at low angles followed by a fast decrease until a 2$\theta$ value of $\sim$ 15$^\circ$. This decrease is followed by a broad peak like feature at $\sim$ 2$\theta$ = 30$^\circ$. Kanai et al. \cite{kanai} studied the effect of mechanical grinding time on the superconducting phases of the Bi-Pb-Sr-Ca-Cu-O system and found the superconducting phases to amorphise with an increase in the grinding time due to breaking of the weak bonds between the Bi-O layers arising due to van der Waals force \cite{suono}. With an increase in the grinding time from 17 minutes to 1 hour, the PXRD spectrum showed a fast decrease of intensity from 3$^\circ$ till 15$^\circ$ followed by a broad hump at 30$^\circ$, exactly similar to our observations. A similar observation was also made by Kim et al. \cite{kim} and Luo et al. \cite{luo2} where they measured the effect of mechanical grinding on the superconducting properties of Ag/Bi-2223 tapes and Bi-2212 powders, respectively. The measured PXRD spectrum till 2$\theta$ $\sim$ 60$^\circ$ showed a similar hump feature at 30$^\circ$. Hence, it is clear that the broad hump seen in the powder diffractograms of Fig. \ref{fig:background} for a large grinding time of 2 hours is due to amorphisation of the Bi-2212 as well as Bi-2201 phase.\\
In order to reduce the amorphisation of the main as well as co-crystallising phases, we made another set of powders with a reduced grinding time of 2 minutes. Green, wine and royal-blue solid curves in Fig. \ref{fig:background} (a), (b) and (c) show the powder diffractograms for RG1, RG2 and PT powders obtained for a grinding time of 2 min respectively. As is immediately apparent from the Fig. \ref{fig:background}, the broad peak at 30$^\circ$ has vanished, suggesting that lower grinding time of 2 minutes ensures that the weak bonds between the Bi-O layers do not break and hence amorphisation of Bi-2212 as well as Bi-2201 is avoided. However, the initial fast decrease of intensity till 20$^\circ$ still occurs. In order to understand the source for this, we performed an empty sample holder run. The resultant powder diffractogram is shown in Fig. \ref{fig:background} (d) which clearly shows a decreasing intensity till 20$^\circ$, after which it becomes constant. For the powder diffractograms that were obtained at a low grinding time of 2 minutes, the amount of sample was quite less, else in a short grinding time of 2 minutes, complete grinding of the powder wouldn't happen. So, it is natural that such a small amount sample would have contributions from the sample holder, as observed. Additionally, since the grinding time is reduced, a large number of crystallites may be produced which suffer from preferred orientation. We performed Rietveld refinement on each of the six curves displayed in Fig. \ref{fig:background} (a)-(c) using JANA and FullPROF softwares, as described below.
\subsection{Profile refinement}
With the parameters of the instrument profile known, we then went ahead to calculate the sample's contribution to the diffraction peak profile. As expected, the instrumental contribution to the diffraction profile was not sufficient to describe the observed profile completely.
\begin{figure}[H]
	\centering
	\includegraphics[width=1\linewidth]{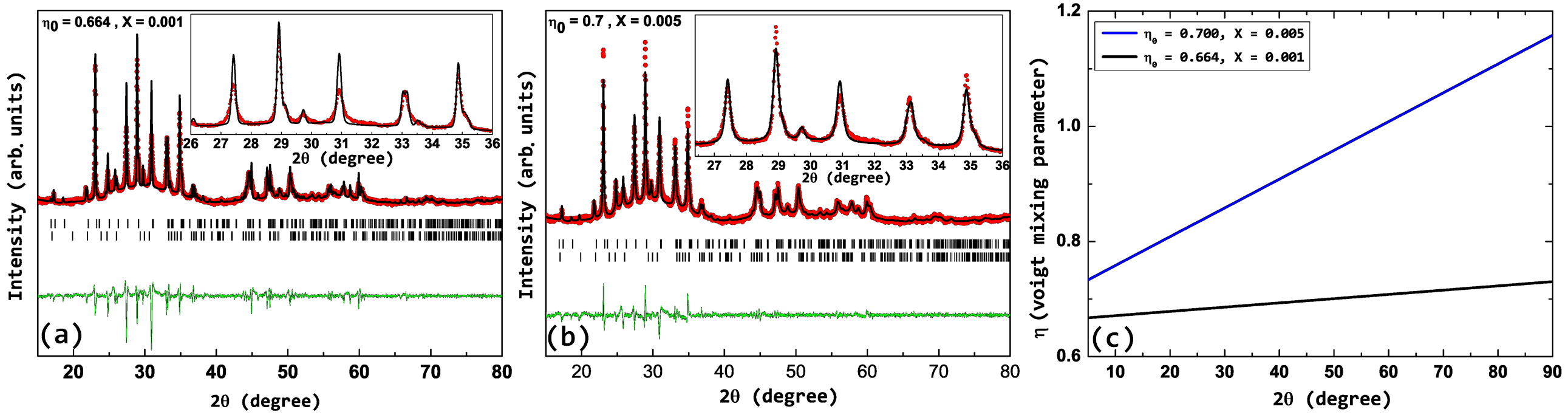}
	\caption{(colour online) (a) Rietveld plot of the PT sample in the 2$\theta$ range of 15$^\circ$ to 80$^\circ$ with profile parameters $U = 0.05613, V = -0.052246, W = 0.029960, \eta_0 = 0.664, X = 0.001$. (b) Rietveld plot of the same PT sample in the 2$\theta$ range of 15$^\circ$ to 80$^\circ$ with profile parameters $U = 0.05613, V = -0.022465, W = 0.029960, \eta_0 = 0.7, X = 0.005$. (c) Plot of the Voigt mixing parameter as a function of 2$\theta$ where black filled circles correspond to $\eta_0$ = 0.664 and $X$ = 0.001 while blue filled circles to $\eta_0$ = 0.7 and X = 0.005.}
	\label{fig:profile-refinement}
\end{figure}
Figure \ref{fig:profile-refinement} (a) shows a Rietveld fit obtained using FullPROF software for the PT-2hour powder sample with the obtained profile parameters. It can be clearly seen that for the peaks corresponding to 2$\theta$ values of 27.4139 (\textit{hkl} $=$ 115 (Bi-2212)), 28.9177 (\textit{hkl} $=$00\textbf{10} (Bi-2212)), 29.7267 (\textit{hkl} $=$ 151 (Bi-2201)) and 30.9318 (\textit{hkl} $=$ 117 (Bi-2212)), the calculated pattern does not match the observed pattern very well. The peaks do not fit the Gaussian and Lorenzian width (FWHM). This kind of $2\theta$ dependence of the FWHM's deviation from the observed profile can be a result of a number of factors, specially in layered materials with intergrowth defects \cite{manaila}. In case of BSCCO, it has been studied by Manaila et al. \cite{manaila}, Malis et al. \cite{malis} and Kulakov et al. \cite{kulakov}, that the observed difference in the FWHM width from the calculated profile may be due to the intergrowth behaviour in the BSCCO system, Cowley short range order parameter and the lattice constants of the intergrowth and host phase. Manaila et al. \cite{manaila} use the Hendrix and Teller theory \cite{hendrix} to describe the peak width and peak shift as a function of defect density of the intergrowth phase and Cowley short range order parameter \cite{cowley}, while Kulakov et al. \cite{kulakov} use the theory by Suzuki et al. \cite{suzuki} to argue that the intergrowth defects can induce width oscillation as well as peak shift in x-ray diffractograms of BSCCO powders. If such defects are not the source of the mismatch observed above, then a refinement with varying the profile parameters should not result in a better agreement between observed and calculated profiles. In order to check this, we took the starting value of the profile parameter as described in the section 4.2.1 and refined it again for the diffractogram obtained on the BSCCO sample by keeping the Caglioti parameters U, V, W as fixed and only varying the profile parameters $\eta_0$ and X. The result is shown in Fig. \ref{fig:profile-refinement} (b), which shows a fit to the peak profile done in this manner. It can be seen that the calculated fit to the observed profile is better when compared to Fig. \ref{fig:profile-refinement} (a). 
However, peaks at 2$\theta$ values of $\sim$ 28.9177 (\textit{hkl} $=$ 0 0 10 (Bi-2212)), 30.9318 (\textit{hkl} $=$ 1 1 7 (Bi-2212)), 34.88 (\textit{hkl} $=$ 0 0 12 (Bi-2212) still have a considerable mismatch between in their respective FWHM's, whereas peaks corresponding to 2$\theta$ value 33.06 ((\textit{hkl} $=$ 2 0 0 (Bi-2212))), 27.4139 ((\textit{hkl} $=$ 1 1 5 (Bi-2201))) show much less deviation in FWHM. This is not typical of a size, strain or shape like broadening as these affects cannot be accounted for by using the pseudo-Voigt approximation \cite{david,thompson}. We also did not find selective peak broadening, i.e. broadening of a family of certain hkl's, occuring possibly,5 due to above mentioned reasons. However, for a quantitative comparison of the nature of the defects and their contribution to the crystal quality, Hendrix and Teller's theory is better suited. This is being done right now and would be published separately.\\
We found that refining the profile parameters over a large 2$\theta$ results in problems. Fig. \ref{fig:profile-refinement} (c) shows a plot of the variation of the Voigt mixing parameter $\eta$ over the 2$\theta$ range of 12- 90 degree, where the black solid curve corresponds to the mixing parameter obtained from the instrument's contribution while the blue curve is that obtained by fitting to the sample powder diffractogram. It can be seen that while $\eta$ corresponding to the instrument's profile, remains below the allowed upper limit of 1 (refer equation \ref{eq:pseuo-voigt}) for all angles, the one corresponding to the sample fit (blue curve) goes beyond 1 at $\sim$ 60$^\circ$! This is not an acceptable scenario since it implies that the new profile parameter values shall start fitting even unidentified peaks, if any, and even though the fit may be better, it would be physically wrong.
\subsubsection{Structure of the main Bi$_{2}$Sr$_{2}$CaCu$_{2}$O$_{8+x}$ (Bi-2212) phase}
So, for all further refinements, we fixed the profile parameters U, V, W, $\eta_0$ and X for FullPROF and GU,GV,GW, L$_x$ and L$_y$ for JANA program. After fixing this, to proceed further with the refinement, the structure of the main Bi-2212 phase as well as the inter-growth phase Bi-2201 and co-crystallisinng Phase$\#$3 was needed. Structure of Bi$_{2}$Sr$_{2}$CaCu$_{2}$O$_{8+x}$ (Bi-2212) phase has been described both in centrosymmetric (Bbmb) as well as non-centrosymmetric (Bb2b, Cc) space groups settings based on x-ray and neutron diffraction measurements performed on single crystals \cite{petricek,kan,beskrovnyi,miles}. Petricek et al. \cite{petricek,petricekjana} report that a non-centrosymmetric description leads to significantly better results, owing to an improved description of oxygen position and its coordination to the Bismuth atom.  However, in such a model a large set of parameters would have to be refined, since removal of the centre of symmetry doubles the number of parameters in the asymmetric unit. Additionally, the two sets of parameters created due to removal of symmetry tend to have large correlations \cite{mironov}. 
\begin{figure}[h]
	\centering
	\includegraphics[width=0.9\linewidth]{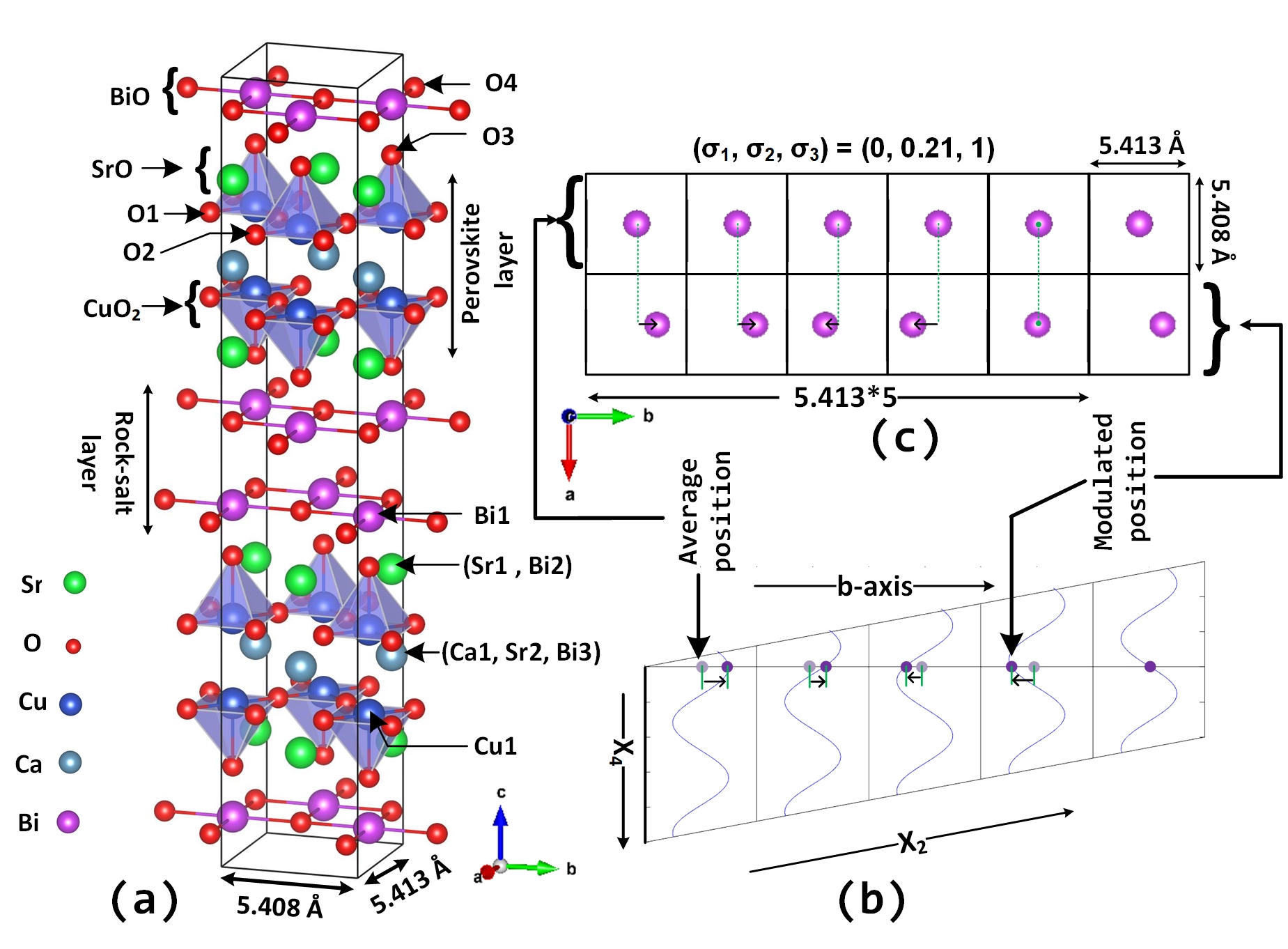}
	\caption{(Colour online) Schematic of the average structure of the Bi$_2$Sr$_2$CaCu$_2$O$_8$ phase. (b) Modulated structure simulated using JANA with modulation vector (0,$\sigma$,1) = (0,0.21,1) and first harmonic sin component for Bi1 atom equal to 0.1 (c) Same modulation represented in x$_2$-x$_4$ section of the superspace.}
	\label{fig:2212-structure}
\end{figure}
Since our powders diffractrograms have smaller number of peaks, we chose not to use Petricek's non-centrosymmetric model for the starting parameters. For the same reason we also did not consider the structure model with a monoclinic space group Cc (non-centrosymmetric) reported by Gladyshevskii et al. \cite{gladyshevskii}, wherein, the structure was first refined in an orthorhombic subcell in space group Ccc2 after which the superstructure was refined in a monoclinic cell (9 times that of the orthorhombic cell), refining 579 parameters with 7227 independent reflections. For these reasons, we chose a centrosymmetric starting model in the space group Bbmb, as fewer parameters are required to describe the structure. Within the centrosymmetric models, we found that models without cation substitution do not fit our data, nor do models with extra oxygen atoms \cite{yamamoto}. The best fit was obtained from the starting structure of Petricek \cite{petricek} et al.\\
The optimised structure obtained is shown in Fig. \ref{fig:2212-structure} (a). In this structure, Bismuth (Bi) atom occupies three different sites: (1) with site symmetry m and occupancy of 0.99. This Bi atom is labeled as Bi1; (2) with site symmetry m and partial occupancy of 0.05 at Sr1 site. This Bi atom is labeled as Bi2; and (3) with site symmetry 2/m and partial occupancy of 0.05 at Sr2 site. This Bi atom is labeled as Bi3. The Bi1 atom also forms the Bi-O layer, where the oxygen atom sits in the position (0.02, 0.157, 0.056). The nomenclature for all the other atoms is followed similarly: atom at a position with maximum occupancy is named as atom1, while it is labeled as atom2/atom3/atom4 where it has a partial occupancy. Sr and Ca atom sites have partial Bismuth substitution of about 5$\%$ each. Oxygen atom's position and occupancy was not refined at all since it is difficult to determine the position of a light atom like oxygen from x-ray diffraction due to the small scattering cross-section of oxygen \cite{petricek,gao,miles}. Additionally, in BSCCO systems, oxygen atoms are surrounded by metal atoms which have high scattering factors compared to oxygen which leads to a further lowering of the visibility of the oxygen atoms. Petricek et al. \cite{petricek} have also shown that x-ray diffraction is not very sensitive to site occupancies of cations, wherein, they found that the R factors before and after refinement of site occupancies remained the same. Therefore, we chose not to refine the site occupancy of any atom. Additionally, there are symmetry constraints for certain atoms that have to be respected. For instance, calcium atom has symmetrically constrained x, y and z coordinates in reports \cite{petricek,miles,yamamoto,kan,beskrovnyi}. Similarly, the y-coordinate for Bi, Sr and Cu atoms are fixed by symmetry. We also did not refine copper's position. Thus, the 4 parameters that we refined with the present data are x and z positions of heaviest atoms Bi1 and Sr1. Table \ref{table:lattice-constant} gives the details of refined lattice constants a, b and c, and Bi1, Sr1 atoms position's refinement.\\ 
From Fig. \ref{fig:2212-structure} (a), it can be seen that the structure comprises perovskite blocks made of layers of CuO$_2$ and SrO between which Ca atoms are sandwiched. Copper occupies position with site symmetry group m with the coordinate (0.5, 0.2499, 0.1965) with an occupation value of 1. This copper atom bonds with the oxygen atom (labelled O2) to form the copper oxide plane. The other block is made of rock-salt comprising Bi-O layers. Bismuth (Bi1) occupies the site 16m with an occupancy of 0.9. Bi1 and the oxygen atom (labelled O4) forms the Bismuth oxide plane. 

\begin{table}[H]
	\centering
	\begin{tabular}{|c|c c c| c c c c |c |c |c|}
		\hline 
		
		Sample   & ~a(\AA)~ & b(\AA)~~ & c(\AA)~ & ~~x-Bi1~~ & ~~z-Bi1~~ & ~~x-Sr1~~ & ~~z-Sr1~~ &  ~~q-vector~~ & \makecell{GOF\\(3D)}  & \makecell{GOF\\(3+1D)}    \\ \hline

		RG1-2hour & 5.396 & 5.423  & 30.859    & 0.2327  & 0.0523 & 0.2890   & 0.1395 & 0.277464 & 5.34  &  5.23  \\ \hline 
		RG2-2hour & 5.409 & 5.407  & 30.875    & 0.2129  & 0.0493 & 0.2525   & 0.1408 & 0.2210   & 3.52  &  3.42  \\ \hline 
		PT-2 Hour & 5.405 & 5.414  & 30.852    & 0.2136  & 0.0516 & 0.2792   & 0.1404 & 0.2139   & 2.92  &  2.70  \\ \hline 
		RG1-2min  & 5.416 & 5.412  & 30.847    & 0.2023  & 0.0523 & 0.2711   & 0.1408 & 0.1998   & 5.23  &  5.11  \\ \hline 
		RG2-2min  & 5.425 & 5.401  & 30.900    & 0.2113  & 0.0511 & 0.2793   & 0.1399 & 0.197918 & 4.60  &  4.46  \\ \hline 
		PT-2min   & 5.389 & 5.408  & 30.855    & 0.2116  & 0.0510 & 0.2995   & 0.1399 & 0.221242 & 2.08  &  1.86  \\ \hline 
	\end{tabular} 
	\caption{Optimised values of lattice constants a, b and c; x and z coordinates of Bi1 and Sr1 atoms; b* component of modulation vector obtained from a Le bail fit using JANA Rietveld refinement for the 2$\theta$ range of 15$^\circ$-65$^\circ$ for RG1-2hour, RG2-2hour, PT-2hour, RG1-2min, RG2-2min and PT-2min samples. Goodness of fit obtained with 3D and 3+1D structural model is also shown.}
	\label{table:lattice-constant}
\end{table}
\subsubsection{Effect of Modulation}
\begin{figure}
	\centering
	\includegraphics[width=0.7\linewidth]{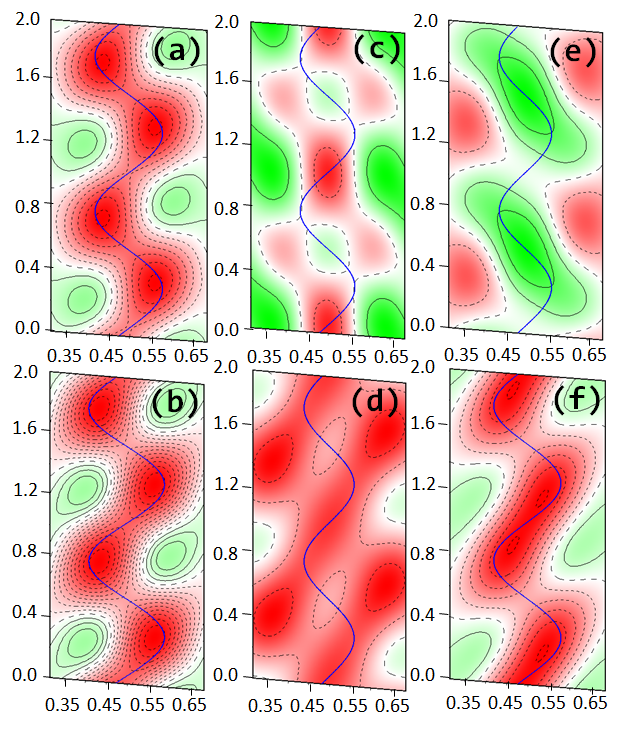}
	\caption{(colour online) $x_{s1}$-$x_{s4}$ section of four dimensional difference Fourier map for Bi-1 atom  for samples at contours of 2 e$\AA^{-3}$. Dash, dots and solid line represent zero, negative and positive density difference respectively. Green and Red colours represent positive and negative electron density differences respectively. 
		(a) RG1-2min ($\rho^{min}$/$\rho^{max}$ $=$-12.41/5.06)  
		(b) RG1-2hour($\rho^{min}$/$\rho^{max}$ $=$-21.49/7.90), 
		(c) RG2-2min($\rho^{min}$/$\rho^{max}$ $=$ -2.90/3.28)
		(d) RG2-2hour ($\rho^{min}$/$\rho^{max}$ $=$-4.89/0.75) 
		(e) PT-2min ($\rho^{min}$/$\rho^{max}$ $=$-10.11/3.19) 
		(f) PT-2hour ($\rho^{min}$/$\rho^{max}$ $=$ -3.95/5.95)}
	\label{fig:difference-map}
\end{figure}
\begin{figure}[H]
	\centering
	\includegraphics[width=0.6\linewidth]{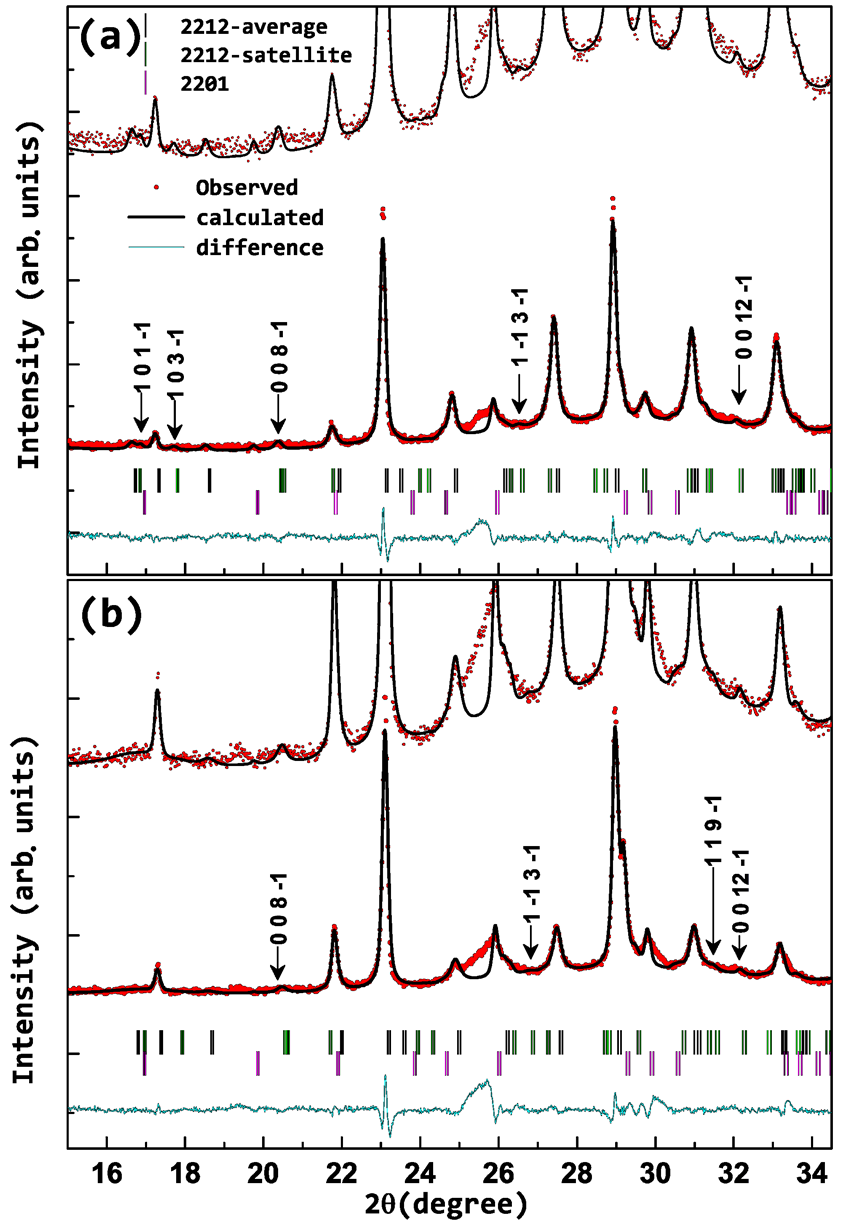}
	\caption{(colour online) Le-bail refinement of the modulated structure in (a) PT-2hour (b) RG2-2hour sample. Red solid circles denote the data points while black solid curve is the Le-bail fit. Satellite reflections are marked by arrows. Vertical bars denote Bragg positions. Blue colour indicates the difference curve.}
	\label{fig:Modulation_lebail}
\end{figure}

The average structure of the Bi-2212 phase that has been described in the above section can not describe all the features of the x-ray diffractograms of Bi-2212. There occur unindexed peaks which could arise from "satellite" reflections arising from modulations in the structure of Bi-2212 \cite{gao,kim,petricek,miles,patterson}. A mismatch between the perovskite block and rock-salt block in Fig. \ref{fig:2212-structure} (a) is conjectured to be the reason for the modulations in the average structure with the modulation vector having a component in the b*-direction \cite{mironov,zhao}. Hence, a 3 + 1D superspace description is needed for explaining the satellite reflections in the powder diffractograms \cite{dewolff,petricekjana2000,yamamoto}. The atomic positions and the site occupancies of the atoms in the modulated structure do not have the translational symmetry in the 3D space. However, in the superspace representation, the translational symmetry is restored, as shown in Fig. \ref{fig:2212-structure} (b) which shows a x$_4$-x$_2$ section, where x$_4$ and x$_2$ represent the directions corresponding to lattice vectors of the direct lattice in the superspace representation (shown in Fig. \ref{fig:2212-structure} b). Lattice vector along x$_4$ is defined such that it is orthogonal to the three dimensional space. A non-modulated structure is shown using light purple atoms and the modulated Bi1 atoms are shown using dark purple colours. The atoms drawn in the dark purple colours do not have the translational symmetry along b-axis but has translational symmetry in the x$_4$-x$_2$ section. Blue wave-like curves represent the superspace translation symmetric counterpart of the point like atoms in dark purple colour. The translation symmetry, along with other allowed symmetries of the modulated structure, form a superspace group in which the structure, and hence, the diffraction pattern must be described. The modulations show characteristic satellite reflections in their diffractograms. We index the satellite reflection using an additional integer "$\textit{m}$" along with the the conventional set of three integers $\textit{h}$,$\textit{k}$ and $\textit{l}$ which are used to label the main reflection of the average or the non-modulated structure \cite{dewolff,petricekjana2000,yamamoto}. The index $\textit{m}$ assumes negative as well as positive integer values.\\
In order to see the distribution of electron density difference of Bi1 atom corresponding to observed modulations and the model used to describe the modulation in the 6 powders, we plotted difference Fourier maps so as to study differences in the electron density in each map. Figures \ref{fig:difference-map} (a)-(f) plot such maps for Bi1 atom in RG1-2min, RG1-2hour, RG2-2min, RG2-2hour, PT-2min and PT-2hour samples respectively at an electron density difference of 2e $\AA ^{-3}$. The modulated position of the Bi1 atom is shown as a solid curvy line in each. It is immediately apparent that the Bi1 atom's position modulation difference is largest in RG1 samples compared to the RG2 and PT samples. Furthermore, the difference maps show clear differences in the modulation of Bi1 atom in RG1-2min, RG2-2min and PT-2min powders compared to RG1-2hour, RG2-2hour and PT-2hour powders, wherein, the modulation difference is found to be more in each 2-hour powder sample compared to the corresponding 2-min sample.\\
It is known that in a Le Bail fit \cite{lebail}, the peak intensities are optimised without the constraints of a structural model, so we first did a Le Bail fit for all the powders. Fig. \ref{fig:Modulation_lebail} shows such a representative Le Bail fit for 2 powders, namely, PT-2 hour (\ref{fig:Modulation_lebail} (a)) and RG2-2 hour (\ref{fig:Modulation_lebail} (b)) using the lattice constants a,b,c and modulation vector components ($q_i$, $q_r$) (explained below) as the fit parameters. From the graph, we find that the XRD peaks do not have a very good resolution (how precisely can a peak be distinguished from the background in a given x-ray diffractogram) with respect to the satellite peaks that arise due to structural modulation. The peaks corresponding to the average structure of the contributing phases have very high intensities, whereas for the satellite peak, only few reflections of the order $|$m$| =$ 1 can be observed.
Top figures in each panel (a) and (b) of Fig. \ref{fig:Modulation_lebail} show a blow-up of the entire data set, shown in the lower figures, to zoom on the satellite peaks. Peak position corresponding to the average structure (without modulation) and those corresponding to the modulated structure (satellite reflections) have been marked by black and green bars. Few of the satellite peaks have been indexed in figure \ref{fig:Modulation_lebail}, where the first three indices are the Miller indices corresponding to the average structure, denoted by $\textit{h}$,$\textit{k}$,$\textit{l}$ respectively, while the last index denoted by $\textit{m}$, represents the satellite reflection \cite{kan,miles,yamamoto,petricek,beskrovnyi}. The obtained q vectors for all the 6 powders are shown in column 9 of Table \ref{table:lattice-constant} above. Exactly similar to the observations made from the electron density maps, we find that (i) the modulation vector q has the highest value for the RG1 powders (both 2 minutes as well as 2 hours) and (ii) the value of modulation vector for a given kind of powder is higher for the 2 hour grinding time compared to the corresponding 2 minutes grinding. Additionally, the q vector decreases systematically from RG1-2hour powder to PT-2 hour powder with an anamolous increase in the PT-2 min powder that could be due to the very high background in the PT-2min powder \cite{toby}. It can also be observed that the goodness of fit reduces in all cases with modulation compared to that without modulation (c.f. columns 10 and 11 of Table \ref{table:lattice-constant}).\\
To do Rietveld refinement, the modulated structure of Bi-2212 phase was described in the superspace group (N Bbmb/1$\bar{1}$1 (Bbmb(0$\gamma$1)) \cite{petricek,miles}. The symbol (0$\gamma$1) represents the 3 components of the modulation wave vector (\textbf{$q$}) along a*, b* and c*. To fit the structural model in the 3 + 1D superspace, the starting point was to use an approximate starting value of the modulation wave vector (\textbf{$q$}), which was obtained from Petricek et al. \cite{petricek}. The modulation wave vector, \textbf{$q$} consists of a rational ($q_r$) and an irrational component q$_i$, which is approximated to the nearest rational value. Accordingly, the starting value of $q$  was taken as (0,0.21,1), where 0.21 is the irrational component and 1 is the rational component. The approximate value of the modulation amplitudes observed for different atoms in the Bi-2212 system was obtained from \cite{petricek,miles}. There are two kinds of modulation that are known to occur in the Bi-2212 system: (1) position modulation and (2) occupation modulation. As a result  of the discussions above, we have not included any occupation modulation and only incorporated the position modulation of Bi1 atom in the present study. The occupancy was fixed at the value determined from the section above. Since the irrational component q$_i$ has the value 0.21, it takes $\sim$ 5 unit cells (4.76 unit cells) for the incommensurate structure to coincide with the approximate commensurate structure, as shown in Figs. \ref{fig:2212-structure} (b) and (c).\\
It was found that except for the RG1 samples and RG2-2min sample where there are un-indexed impurity phases and a third phase making the assignment of satellite peaks impossible, we were clearly able to observe and assign satellite peaks corresponding to PT-2hour and RG2-2hour samples which suffer minimum intergrowth problems. The PT-2min powder sample also did not have any impurity peak but a fit incorporating the modulations was not possible in this powder sample because of its high background at lower angles. A Rietveld fit incorporating the effects of modulations is shown in Figs. \ref{fig:Rietveld-Refinement} (d) and (f).
\begin{figure}
	\centering
	\includegraphics[width=0.8\linewidth]{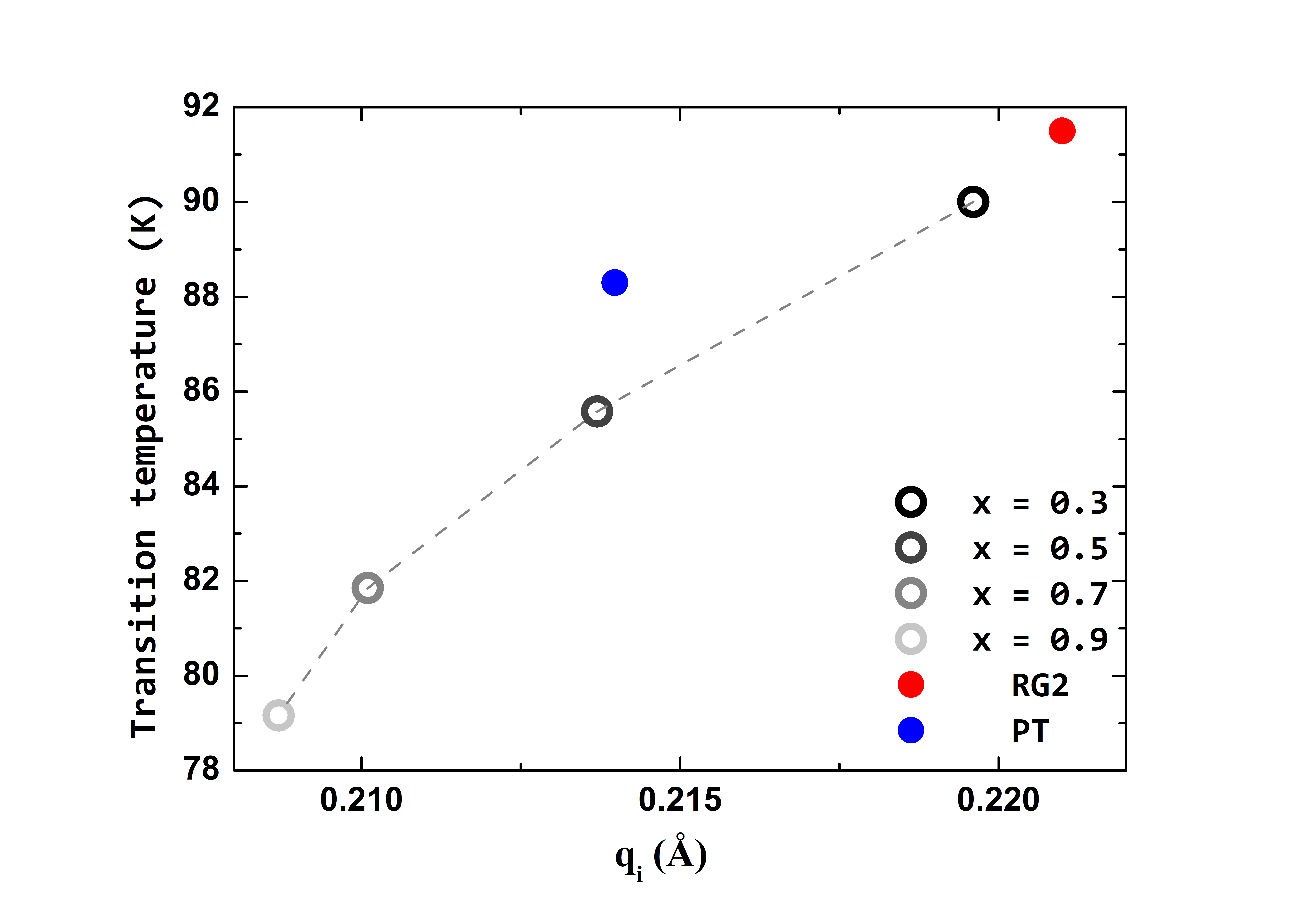}
	\caption{(Colour online) Filled solid circles represent the variation of the b* component of the modulation vector $q_i$ vs. transition temperature T$_c$ for RG2-2hour and PT-2hour samples. Open circles depict the $q_i$ variation with T$_c$ of Bi$_2$Sr$_{2-x}$Ca$_x$O$_{6+\delta}$ \cite{zhou} drawn for comparison.}
	\label{fig:modulation-SC}
\end{figure}
Zhou et al. \cite{zhou} studied the effect of substitution of Ca atom on the Sr site in the 2201 superconductor, Bi$_2$Sr$_{2-x}$Ca$_x$CuO$_{6+\delta}$, and the correlation between the modulation wave-vectors and the transition temperature T$_c$. They found that as the amount of impurity, namely Ca content, increases, the modulation vector q$_i$ increases and the superconducting transition temperature T$_c$ also increases. They explained this observation by a crystal mismatch model where the mismatch between the perovskite CuO$_2$ layer block and rock-salt Bi-O layer block causes the modulation in the 2201 superconductor, due to an increased corrugation of CuO$_2$ layers. In order to see if a similar correlation exists between our samples, we plotted the modulation vector q$_i$ obtained from Rietved refinement of RG2-2hour and PT-2hour powders (see Fig. \ref{fig:Rietveld-Refinement}), and the superconducting transition temperature T$_c$ \cite{rajak}, as shown in Fig. \ref{fig:modulation-SC}. For comparison, q$_i$ and T$_c$ variation obtained by Le-bail refinement in Ca substituted Bi-2201 \cite{zhou} is shown by open grey circles. It can be observed that the magnitude of modulation vector in our Bi-2212 system is similar to that of Bi-2201 \cite{zhou}. Additionally, the modulation vector q$_i$ along with T$_c$, has a higher value for the RG2-2hour sample than the PT-2hour sample. This observation, then, suggests that the mismatch between the CuO$_2$ planes and Bi-O planes is lesser in the PT samples and higher in the RG2 samples. From the quantitative phase analysis obtained in Table \ref{table:final fit values}, it was found that PT samples had a higher phase fraction of BSCCO-2212 phase and no impurity phase as compared to the RG2-2hour samples. So, a lower superconducting transition temperature in PT samples was counter-intutive since lesser disorder always results in a higher transition temperature \cite{landau}. However, in the special case of BSCCO superconductor, a decrease in the superconducting transition temperature, infact, is a pointer to better crystal quality since it implies a lower mismatch between the CuO$_2$ and Bi-O layers, and consequently, a lower corrugation of the CuO$_2$ planes. Since the powder samples are prepared from crushing single crystals, this observation, then, means that the single crystals grown from the PT technique are better in quality than the crystals grown using the regrowth technique.
\subsection{Texture refinement}    
Rietveld refinement with incorporation of the structure alone didn't give a very good fit and we observed a large deviation of the intensity from the theoretically calculated intensities. It is very well known that Bi-2212 system suffers from preferred orientation along the $\{$00\textit{l}$\}$ direction \cite{schmahl,dellicour,zhao,rajak}. This effect is expected to be enhanced in our powders since they are obtained by grinding a large number of single crystals, and hence, are expected to be textured. So, many grains of the powders are expected to be preferentially oriented along a particular crystallographic {hkl} direction rather than be completely randomly oriented. Moreover, the loading and pressing of the powders on to the sample holder of the x-ray diffractometer introduces additional inconsistency in the obtained pattern making it impossible to get a reproducible data for the same powder on the same diffractometer for different mounts of the sample. To incorporate the effect of the preferred orientation, we refined the preferred orientation parameters G1 and G2 of the preferred orientation function, in an attempt to improve the fit. As described in section 3, the preferred orientation correction was accomplished using the March-Dollase function \cite{dollase}, given by $P_{h,\phi}$ (modified march equation in FullPROF software\cite{carvajal,carvajal1}), which describes the modulations in $I_{h,\phi}$ due to a non random distribution of crystallite orientation. G1 and G2 can be refined in either JANA or FullPROF software \cite{carvajal,carvajal1,petricekjana}.
\begin{figure}[h]
	\centering
	\includegraphics[width=1\linewidth]{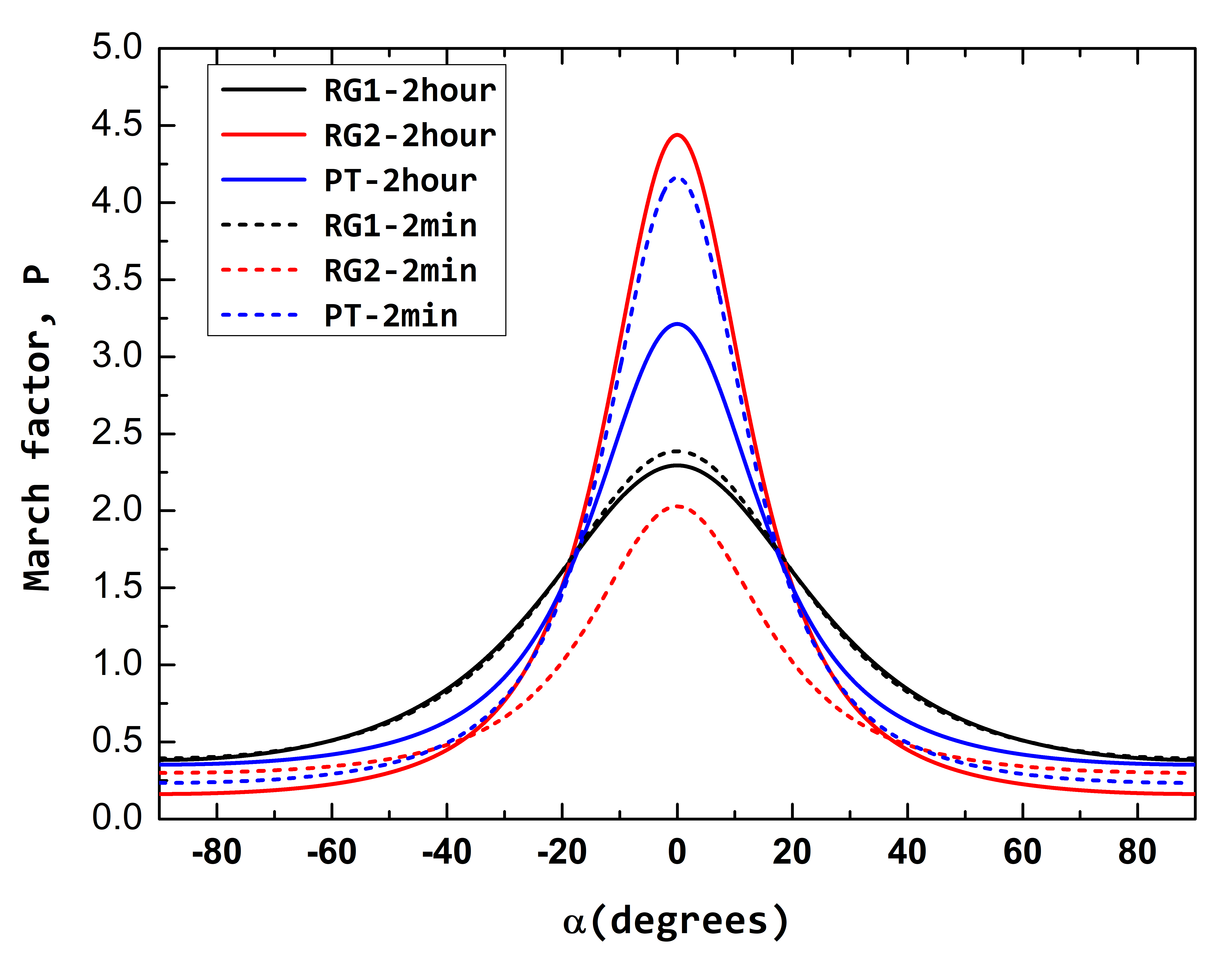}
	\caption{(Colour online) Plot of the variation of the March factor, P, with respect to the angle $\alpha$ (see text for details). Solid curves denote the preferred orientation function for RG1-2 hour (black), RG2-2 hour (red) and PT-2 hour (blue) samples. Dotted curves with the same colour code denote the March parameter for lower grinding time of 2 minutes.}
	\label{fig:marchplot}
\end{figure}
Figure \ref{fig:marchplot} plots the March parameter, \textit{P}, with respect to acute angle $\alpha$ between the normal to the diffracting plane (hkl) and the platelet normal (00l). It is expected that the powders that have been made using a small grinding time of 2 minutes will suffer from a higher preferred orientation since not enough grains with random orientations could be produced. This effect can be seen in powders of RG1 and PT, where the preferred orientation value for RG1-2 min and PT-2 min is higher than the corresponding values for RG1-2 hour and PT-2 hour respectively. However, the effect is seen to be reversed in RG2 powders where the preferred orientation factor is higher for RG2-2 hour powder than RG2-2 min powder. We suspect that it may be due to effects associated with sample mounting where the powders need to be gently pressed before data acquisition to avoid sample height/displacement error, otherwise an error in shift of peak positions may occur \cite{parrish}. Such mountings also lead to added preferred orientations in the powders and it is never possible to completely eliminate this effect.\\  
Values of G$_{1}$ and G$_{2}$ obtained from fitting the preferred orientation function $P_{h,\phi}$ for various powders are summarised in Table \ref{table:final fit values}. It can be seen that for all the samples with either the grinding time of 2 hours or of 2 minutes, the values of G$_{1}$ is less than 1 and greater than 0, indicating that the grains in the powder sample have a platy habit \cite{carvajal,carvajal1}. This is expected since the powders were obtained by grinding the single crystals which have layered morphology, so the resultant ground powder is also expected to retain the morphology. From Table \ref{table:final fit values}, it is also clear that the RG1 powders have the highest preferred orientation compared to RG2 powders or PT powders. Additionally, G2 which represents the fraction of the untextured powder \cite{carvajal,carvajal1}, is found to be zero for RG1-2hour powder and very small for RG1-2min powder, suggesting that the RG1 powders have virtually very few grains that are untextured.
\subsection{Absorption correction}
\begin{figure}
	\begin{center}
		\includegraphics[width=0.80\textwidth]{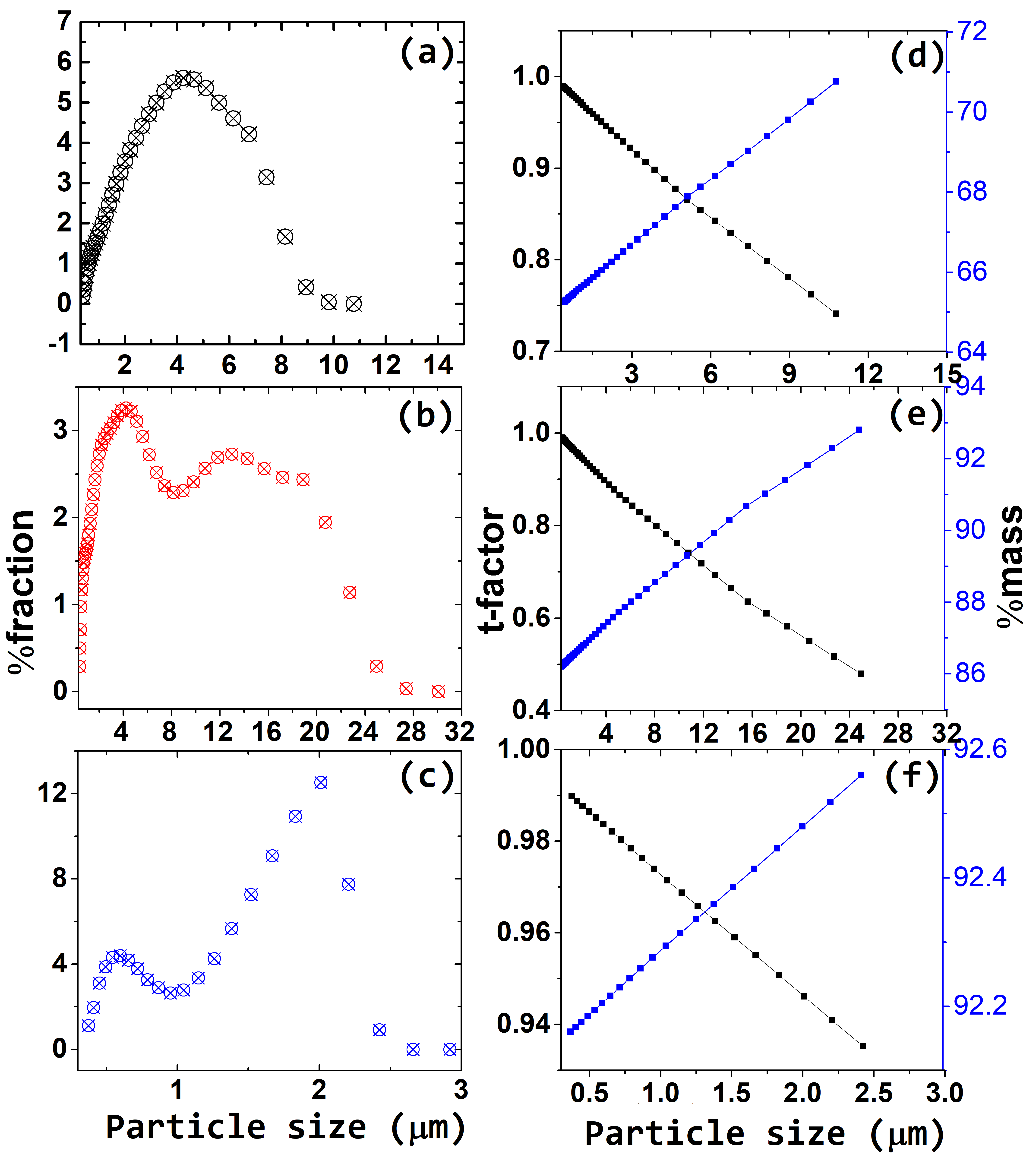}
		\caption{(colour online) Figs. (a), (b) and (c) show the particle size distribution for the RG1-2hour, RG2-2hour and PT-2hour powder samples respectively. Black filled circles in Figs. (d), (e) and (f)  show the variation of t-factor for the Bi-2212 phase for different average particle size of the grain of the powder while the blue filled circles show the variation of the estimated phase fraction of the Bi-2212 phase as a function of the average particle size.}
		\label{fig:brindley-factor}
	\end{center}
\end{figure}

As mentioned in section 3, the absorption coefficient, $\mu_{\phi}$, of the main Bi-2212 phase and the intergrowth phase Bi-2201 vary by $\sim$ 200 cm$^{-1}$, so it is necessary to incorporate absorption correction in the analysis, else the more absorbing phase gets underestimated \cite{monecke,brindley}. The effects of microabsorption were corrected in the Rietveld refinement by incorporating the absorption correction parameter $t_{\phi,h}$. $t_{\phi}$ has been tabulated by Brindley as a function of ($\mu_{\phi}$ - $\overline{\mu}$)D \cite{brindley} vs. $\theta$, where $\mu_{\phi}$ is the phase under consideration, $\overline{\mu}$ is the average linear absorption coefficient for the powder and D is the mean-grain size of the phase. For the Bi-2212 phase, $t_{\phi}$ has been assigned a value 1. This can be done since for every other phase $\phi$, $t_{\phi}$ will be calculated with respect to $t_{\phi}$ for the Bi-2212 phase \cite{schmahl}. To do this, the mean-particle (grain) size, D, of each constituent must be known. Figs. \ref{fig:brindley-factor} (a), (b) and (c) plot the percentage fraction variation with size of the RG1-2hour, RG2-2hour and PT-2hour powders respectively. It can be observed that while the RG1-2hour powder has a uni-modal distribution, both the RG2-2 hour as well as PT-2 hour powders have a bi-modal distribution. Hence, an average particle size may be determined for RG1-2hour powder, but it is difficult to assign an average particle size for the PT-2hour powder. This means that we cannot assign a single value of D to calculate the $t_{\phi}$ for the 2201 phase. However, it can be seen that the size variation in PT-2hour powder is not much: it varies only between 0 and 3. On the other hand, the size variation in the RG-2hour powder is rather large from 0 to 32 $\mu$m. 
To see how the particle size variation effects the estimation of 2212 phase for the three powders, we calculated the variation in t-factor and the estimated phase fraction of the Bi-2212 phase as a function of different average particle size. The results are shown in Figs. \ref{fig:brindley-factor} (d)-(f). It can be seen that while the variation in the estimated 2212 phase for PT-2hour powder is extremely small for the bi-modal size distribution, that of RG2-2hour is larger, in comparison. So, the narrower is the particle size distribution with uni-modality, the better is the phase estimation.
\begin{figure}[H]
	\begin{adjustwidth}{-0.5in}{0.2in}
		\includegraphics[width=1.1\linewidth]{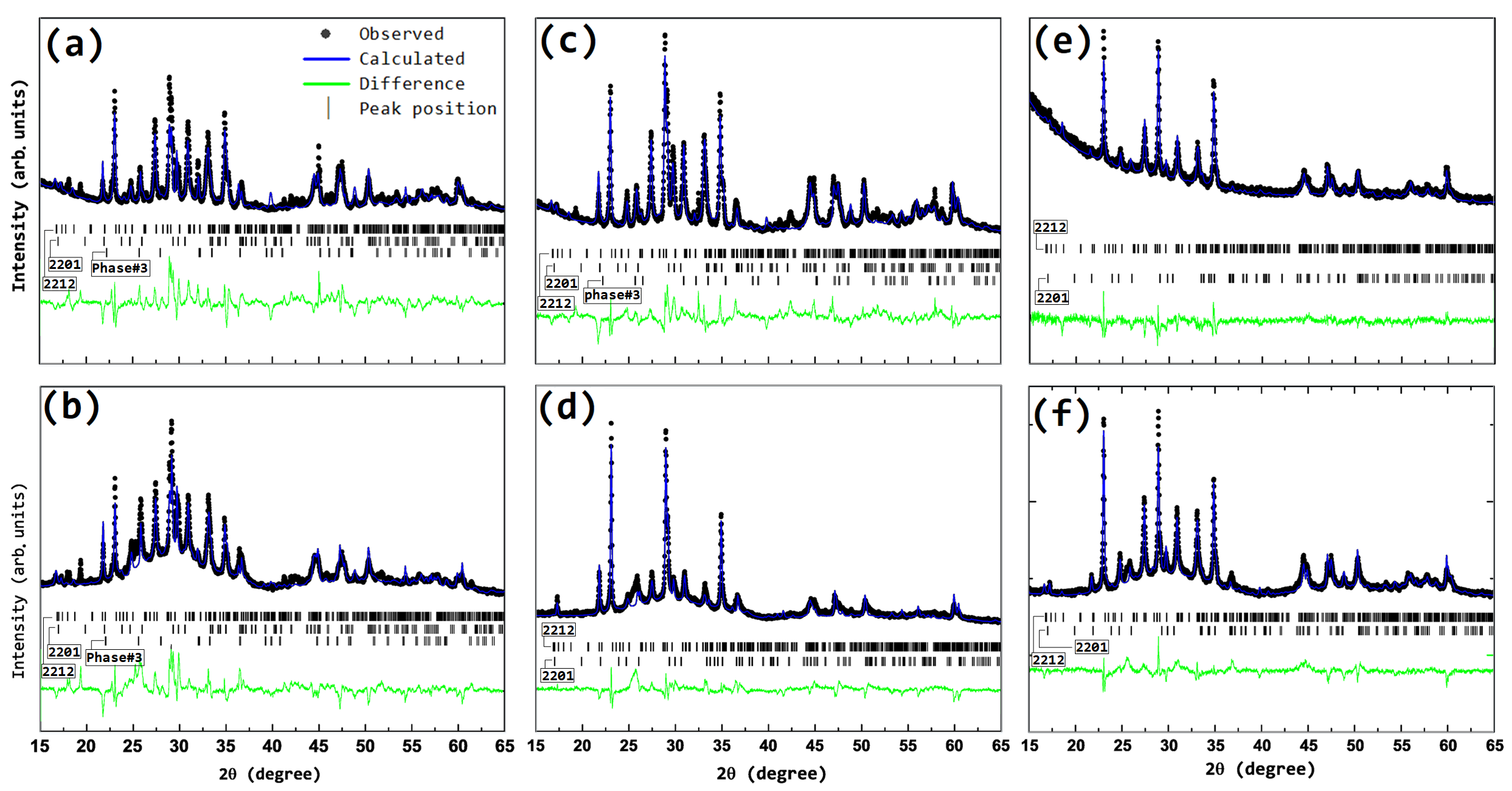}
	\end{adjustwidth}
	\caption{(colour online) Rietveld refinement fit of the diffractogram obtained on the samples (a) RG1-2min, (b) RG1-2hour, (c) RG2-2min, (d) RG2-2hour, (e) PT-2min and (f) PT-2hour. Red filled circles correspond to the observed intensity, black line is the fit to the corresponding data, blue is the difference curve and black bars is the Bragg positions of the main and the first order satellite reflections.}
	\label{fig:Rietveld-Refinement}
\end{figure}
\subsection{Final Rietveld fits}
Figures \ref{fig:Rietveld-Refinement} (a)-(f) show a Rietveld fit to the powder diffraction data of RG1-2min, RG1-2hour, RG2-2min, RG2-2hour, PT-2min and PT-2hour powders respectively. The refinement was done by incorporating the average structure refinement of the main Bi-2212 phase, wherein, the lattice constants a, b and c was refined along with x and z position refinement of the heavy Bi and Sr atoms (see Table \ref{table:lattice-constant}); Bi atoms modulation in the a-direction; lattice constants of the intergrowth Bi-2201 phase and co-crystallising Phase $\#$3; texture refinement and absorption correction refinement.\\
Black filled circles correspond to the data points while blue solid line is a Rietveld fit obtained using JANA software. The results obtained using the FullPROF software are exactly the same and are not shown for brevity. The experimentally obtained diffractograms are characterised by their flat backgrounds for angles greater than 20$^{\circ}$ for 2 minute grinding time, as discussed above. Bragg positions corresponding to Bi-2212, Bi-2201 and Phase $\#$3 have been marked by black bars in all the 6 graphs. It can be seen that the Phase$\#$3 is present in RG1-2min, RG1-2hour and RG2-2min samples (see Figs. \ref{fig:Rietveld-Refinement} (a)-(c)). Since the RG2-2hour sample is made by grinding the single crystals grown by regrowth technique, the presence of Phase$\#$3 in a 2 minute grinding time but its absence in a 2-hour grinding time points to the fact that the Phase$\#$3 is always present in crystals grown by the regrowth technique and its presence in a batch of powder made by grinding the crystals is just statistical. However, for the crystals grown by pressure technique, the Phase$\#$3 was found as a separate phase during the crystal growth process in many crystal growth runs which could be segregated, thus, reducing the co-crystallising phase in the PT crystals. Values of the various refined parameters are tabulated in Tables \ref{table:lattice-constant} and \ref{table:final fit values}.
\begin{table}[H]
	\centering
	\begin{tabular}{| c | c | c  c | c c c | c |}
		\hline 
		\multirow{2}{*}{Sample} &\multirow{2}{*}{\makecell{Zero\\Correction ($\times$ 0.01)}} &\multicolumn{2}{c|}{March Parameter}& \multicolumn{3}{c|}{Phase fraction}&\multirow{2}{*}{GOF} \\ 
		&                & G1        & G2       &  Bi-2212& Bi-22201& Phase$\#$3 &      \\ \hline 
		RG1-2hour           & 0.782508       & 0.671672  & 0.0      & 68.79   & 28.52   & 2.69       & 5.34 \\ \hline 
		RG2-2hour           &-5.760303       & 0.462362  & 0.053821 & 85.95   & 14.05   & 0.0        & 3.42 \\ \hline 
		PT-2hour            & 0.921756       & 0.512909  & 0.333689 & 93.14   & 6.86    & 0.0        & 2.70 \\ \hline 
		RG1-2min            &0.247914        & 0.647525  & 0.093269 & 61.85   & 15.29   & 22.87      & 5.23 \\ \hline 
		RG2-2min            &-0.176593       & 0.521360  & 0.438814 & 51.78   & 37.76   & 10.46      & 4.60 \\ \hline 
		PT-2min             &7.477808        & 0.460654  & 0.218266 & 93.69   & 6.31    & 0.0        & 2.08 \\ \hline 
	\end{tabular} 
	\caption{Zero correction, Preferred orientation correction parameters G1 and G2, Phase fractions of phases present in each sample and Goodness of fit obtained for RG1-2hour, RG2-2hour, PT-2hour, RG1-2min, RG2-2min and PT-2min samples.}
	\label{table:final fit values}
\end{table}
The first observation to be made from Table \ref{table:final fit values} is that we have been able to get excellent goodness of fits with values $<$ 5.5 for all the powders. For the PT powders, which has the absence of any co-crystallising Phase$\#$ 3, the GoF value is the least at 2.08 for the 2-min powder followed by 2.7 for the 2-hour powder. Considering the complication of the 2212 system, these fit values are exceptional. The second observation is the phase estimation of the main Bi-2212 phase, intergrowth Bi-2201 phase and co-crystallising Phase$\#$3 in each sample. For the PT-2min sample, the amount of Bi-2212 phase is the highest at 93.68$\%$, followed by PT-2hour sample where it is 93.14$\%$. It can be observed that the phase estimation in PT powders is more or less independent of the grinding time. The phase estimation of the main Bi-2212 phase is also found to be independent of the grinding time for RG1 powders, where it is in the range of 60-70$\%$. However, for the RG2 samples, grinding time was found to have a large effect on the phase estimation: it is at 86$\%$ for samples with a high grinding time of 2 hours but is at 52$\%$ for powders ground for 2 minutes. However, it is to be noted that the former powder had no co-crystallising Phase$\#$3 but the latter powder had a co-crystallising Phase$\#$3. So, the phase estimation crucially depends on the presence or absence of a co-crystallising phase rather than the grinding time used. Since the powders are made by crushing single crystals grown using various self-flux techniques, the high value of phase fraction of Bi-2212 in PT powders suggest that the PT crystals are much better in quality compared to those of regrowth ones. This observation is also confirmed by the magnetisation measurements \cite{rajak}, wherein, PT crystals showed a sharper superconducting transition when compared to the regrowth crystals. It was also found that PT crystals have a better expected ideal ratio of Ca/(Ca+Sr) of 0.33, implying a better control on the amount of oxygen in these crystals as compared to RG crystals \cite{rajak}.  Adding to the analysis presented in this work, the following reasons are ascribed to the better quality of PT crystals compared to the regrowth crystals: (1) lack of co-crystallising Phase$\#$3 crystals, (2) better alignment of the rock-salt Bi-O layers compared to the perovskite CuO$_2$ and Sr-O layers in PT crystals and (3) better obtainment of the expected ideal ratio Ca/(Ca+Sr) of 0.33, implying a better control of the amount of oxygen.
\section*{Conclusions:}
In conclusion, we have performed detailed Rietveld refinements on powder diffractograms of Bi$_2$Sr$_2$CaCu$_2$O$_{8+x}$ (Bi-2212), where the powders were made by grinding the single crystals of Bi-2212 with the aim of ascertaining the phase purity of the crystals since they are known to suffer from the problem of intergrowth of Bi-2201 phase as well as competing co-crystallising phases. For the analysis, we made powders using a high grinding time of 2 hours and a low grinding time of 2 minutes. This was done to avoid the breakage of the Bi-O layer that is bound by weak van-der Waal's force. However, this resulted in a high background contribution due to sample holder at lower angles making the assignment of satellite peaks difficult in the 2-min samples. For all other analysis, 2-min samples presented a good data. For the Rietveld analysis, two softwares, namely JANA and FullProf were used both of which gave the same results. We incorporated information not only of the average structure of the main Bi-2212 phase but also the modulation of Bi-atoms in the refinement which could be done only for PT-2hour and RG2-2 hour samples. It was found that the modulation vector q$_i$, along with superconducting transition temperature T$_c$ is higher in the RG2-2hour powder than PT-2hour powder, implying a better alignment of the Bi-O planes compared to the CuO$_2$ planes resulting in a lesser corrugation of CuO$_2$ layers. Since the powders were made by crushing single crystals that grow as platelets, a preferred orientation is expected to be present in the powders. This was incorporated by refining the parameters of the preferred orientation function proposed by March and Dollase and $G_1$ parameter was found to be lower for the lower grinding time of 2 minutes as compared to the higher grinding time of 2 hours in samples of RG1 and PT, representing higher preferred orientation in 2 minutes samples of RG1 and PT as compared to the corresponding samples for a higher grinding time of 2 hour. Finally, to incorporate the absorption correction to the Rietveld refinement, a size variation of the mean-particle size with percentage phase estimation was done, where the maximum phase was found for the PT samples. The final estimation of phase fraction by doing such quantitative phase analysis for found to be highest for the PT samples at $\sim$ 94$\%$ for the PT-2min sample, implying that our simple technique of applying pressure while growing the single crystals of Bi-2212 gives excellent quality single crystals, not only in one single crystal but in a batch of crystals. The study is very important since the information on phase purity of a given single crystal is extremely important for the analysis of its other physical properties.

\section*{Acknowledgement:}
The authors acknowledge R. Suryanarayanan for fruitful scientific discussions. D. J-N acknowledges financial support from SERB-DST, Govt. of India (Grant No. YSS/2015/001743).

\end{document}